\documentclass[floatfix,pdftex,prc, 11pt]{revtex4-1}
\usepackage{color} 
\usepackage{fullpage}
\usepackage{graphicx} 
\usepackage{amssymb} 
\usepackage{amsmath} 
\usepackage[utf8]{inputenc} 

\begin{document}
\title{Viscous Hydrodynamics and Relativistic Heavy Ion Collisions}
\author{Joshua Vredevoogd and Scott Pratt}
\affiliation{Department of Physics and Astronomy and National Superconducting Cyclotron Laboratory\\
Michigan State University\\
East Lansing, Michigan 48824}
\date{\today}

\begin{abstract}
The matter created in relativistic heavy ion collisions is fairly well described by ideal hydrodynamics, and somewhat better described by viscous hydrodynamics. To this point, most viscous calculations have been two-dimensional, based on an assumption of Bjorken boost invariance along the beam axis. Here, first results are presented for a fully three-dimensional viscous model. The model is described and tests of the numerical accuracy of the code are presented. Two- and three-dimensional runs are compared, and modest changes are observed for mid-rapidity observables at the highest RHIC (Relativistic Heavy Ion Collider) energies.
\end{abstract}

\maketitle

\section{Introduction}
Hydrodynamics has been central to the interpretation of experimental results from the Relativistic Heavy Ion Collider (RHIC) \cite{:2008ez, Abelev:2008ed}. Even relatively simple hydrodynamic models that include neither the effects of viscosity, a realistic phase transition, nor hadronic rescattering are able to roughly reproduce such important observables as the transverse momentum spectra and the large anisotropic flow observed at mid-rapidity \cite{Bearden:2004yx}.
Despite difficulties in describing these mid-rapidity observables simultaneously with the femtoscopic source sizes and elliptic flow at non-zero longitudinal rapidity \cite{Hirano:2002hv, Heinz:2002un},
the results strongly suggested underlying collective behavior and were taken as strong evidence for the formation of a Quark-Gluon Plasma (QGP).

Since that time, considerable effort has been expended by the modeling community to characterize and constrain the properties of the matter created in these heavy ion collisions.
In particular, several models have focused on developing viscous hydrodynamics for the detailed description of anisotropic flow at mid-rapidity to constrain the viscosity of the QGP \cite{Romatschke:2009im, Song:2007ux}.
To simplify the calculations, most of these models have restricted themselves to mid-rapidity and used the expected approximate boost invariance near zero rapidity \cite{Bjorken:1982qr} to motivate two-dimensional treatments.

In reality, the spectrum of produced particles is well described by a Gaussian of radius $\approx$1.6 units in rapidity \cite{Bearden:2004yx}, 
and one questions the extent to which the finiteness of the extent affects the evolution at mid-rapidity. 
Furthermore, the character of the results away from mid-rapidity display some interesting trends that likely provide some unique and useful constraints on the character of the matter produced and the important features of the theory that governs the production of that matter.
To this end, we endeavor to explore the predictions of viscous hydrodynamics in the full configuration space.

\section{Hydrodynamic Theory}

We consider a system of interacting particles. When the density of particles is relatively low and the mean free paths for particles in the systems are larger than characteristic inter-particle separations, the system can be described as a series of binary collisions. One can then describe the evolution with a Boltzmann equation, which requires following the evolution of the six-dimensional phase space density, $f({\bf p},{\bf r},t)$, for each species. When the mean free path is much smaller than the characteristic size of the system, the local phase space distribution takes on a thermal form, and the system and its evolution can be described by following a small number of variables in configuration space, i.e., current densities ($\rho^\mu$) and the stress-energy tensor ($T_{\mu \nu}$). The equations of motion for the quantities are driven by the conservation laws:
\begin{eqnarray}
\partial_\mu \rho^\mu(x)&=&0,\\
\partial_\mu T^{\mu\nu}(x)&=&0.
\end{eqnarray}

In terms of thermodynamic variables, i.e. energy density ($\epsilon$) and pressure ($p$), the stress-energy tensor for ideal hydrodynamics is 
\begin{equation}
T^{\mu \nu} = (\epsilon + p) u^\mu u^\nu - g^{\mu \nu} p
\end{equation}
where $g^{\mu \nu}$ = diag(1, -1, -1, -1) is the metric tensor.
This expression makes clear that one still requires an equation of state to link the thermodynamic variables - i.e. $p(\epsilon, \rho)$ - in order to close the system of equations.

If the distribution function is allowed to deviate modestly from the equilibrium value, the theory may be extended to account for the effects of a finite mean free path \cite{Muronga:2001zk, Betz:2008me} with viscous treatments.
Roughly, one assumes that collisions cause the stress-energy tensor to relax toward the Navier-Stokes value or that the corresponding modification to the stress energy tensor ($\pi^{\mu \nu}$) leads to a quadratic increase to the equilibrium entropy density \cite{Baier:2006um}.
The equations of motion are then taken either from higher moments of the Boltzmann equations or from the local conservation of the modified entropy density \cite{ Romatschke:2009kr, Pratt:2007gj,Muronga:2006zx,Muronga:2006zw, Huovinen:2008te}.
Such a construction requires the introduction of several new transport coefficients, but supplies equations of motion that apply to systems further from equilibrium.
In this case, we will only allow the transport coefficients associated with shear viscosity to be non-zero. This introduces two new quantities: the shear viscosity ($\eta$) and the shear relaxation time ($\tau_\eta$).
The resulting equations of motion are the conservation of stress energy tensor and the relaxation of the shear tensor to the Navier-Stokes value \cite{Pratt:2007gj,Muronga:2006zx,Muronga:2006zw}:
\begin{eqnarray}
T^{\mu \nu} &=& (\epsilon +p) u^\mu u^\nu - g^{\mu \nu} p + \pi^{\mu \nu} \\
\tilde{\partial}_t \left(\frac{\tilde{\pi}^{ij}}{\sigma_\eta}\right) &=& - \frac{\tilde{\pi}^{ij} - \tilde{\pi}^{ij}_{NS}}{\sigma_\eta\tau_\eta},\\
\tilde{\pi}_{NS}^{ij}&=&-\eta\left(\tilde{\partial}^i n^j+\tilde{\partial}^j n^i
-(2/3)(\tilde{\partial}\cdot n)\delta_{ij}\right).
\end{eqnarray}
The tildes denote quantities determined in the frame of the matter, and roman indices denote that only spatial indices are being considered. Here, $\pi^{\mu\nu}$ is the viscous correction to the stress-energy tensor.  
In the frame of the matter $\tilde{\pi}^{00}=\tilde{\pi}^{0i}=0$ due its orthogonality to the collective velocity.
The quantity $\sigma_\eta$ is related to the equilibrium fluctuation of $\pi_{ij}$, and has the same units of pressure \cite{Pratt:2007gj}. 
For a conformal theory it can be replaced with the energy density \cite{Romatschke:2009kr}.

For reasons of clarity, we choose to tabulate moments of the local shear tensor defined 
\begin{eqnarray}
a_1 = \frac{1}{2} \left( \tilde{\pi}^{xx} - \tilde{\pi}^{yy}\right);  && ~~~~  a_2 = \frac{1}{\sqrt{12}} \left( \tilde{\pi}^{xx} + \tilde{\pi}^{yy} - 2 \tilde{\pi}^{zz} \right) ; \\
a_3 = \tilde{\pi}^{xy};  &&  a_4 = \tilde{\pi}^{xz} ;  ~~~~~ a_5 = \tilde{\pi}^{yz};  \nonumber \\
 b &=&  (1/3) \left[ \tilde{\pi}^{xx} + \tilde{\pi}^{yy} + \tilde{\pi}^{zz} \right] . \nonumber
\end{eqnarray}

This choice of moments clearly separates the shear effects of the strong initial longitudinal expansion into the variable $a_2$, whereas $a_1$ will contain only the effects of the anisotropic transverse expansion.
Furthermore, we have limited the number of variables to the minimum - gaining speed but sacrificing the possibility of a check on the numerical accuracy of the solution.

Hydrodynamic models best apply to the matter at small spatial rapidity soon after the rapidly receding nuclei have passed through one another. The spatial rapidity is defined by
\begin{equation}
\eta_s = \frac{1}{2} ln \left( \frac{ t+z}{t-z} \right) ; ~~~ \tau = \sqrt{ t^2 - z^2},
\end{equation}
and is a measure of the longitudinal position, i.e., $\delta\eta_s=\delta z/\tau$. For a boost-invariant system, the collective velocity is $v_z=z/t$, which makes the spatial rapidity equal to the rapidity of the matter. The condition $v_z=z/t$ describes a system where each density element had a trajectory that began at $z=t=0$, and did not accelerate from that point. From the perspective of hydrodynamics, the lack of acceleration is expected if intrinsic quantities, such as energy density, vary slowly with $\eta_s$ and depend mainly on the proper time $\tau$. This approximate symmetry motivates using hydrodynamic prescriptions where $\eta_s$ is neglected, even in the presence of viscosity, which then reduces the dimensionality of the model from three to two. Of course, such a model only applies to observables measured near mid-rapidity. A principal goal of this study is to discern the validity of this approximation by comparing two- and three-dimensional versions.

\section{Initial Conditions}

While considerable progress has been made into viscous hydrodynamics for event-by-event conditions with fluctuations \cite{Hirano:2009ah, Holopainen:2010gz, Werner:2010aa}, for this paper
we choose a simpler approach and assume averaged initial conditions such as 
those from the optical Glauber model \cite{Miller:2007ri}.
We begin with a Woods-Saxon distribution of nucleons
\begin{equation}
\rho (r) = \frac{\rho_0}{1 + \exp{[(r-R)/\chi]}}
\end{equation}
where $R= 6.37$ fm, $\chi = 0.54$ fm, and $\rho_0 = 0.17$ fm$^{-3}$ for Au nuclei.
One then considers the total probability of interaction as the nuclei pass completely through one another, which is computed via the nuclear thickness function
\begin{equation}
T(x,y) =  \int_{-\infty}^\infty dz~\rho(x,y,z)
\end{equation}

The initial shape of the system is then scaled either to the number of wounded nucleons or the number of binary collisions.
The difference between the scalings comes from the eligibility for multiple interactions.
In the binary collision picture, nucleons are allowed to interact with one another multiple times during the crossing.
This leads to scaling like the product of the two thickness functions
\begin{equation}
n_{BC} (x,y,b) = \sigma T_A(x +b/2,y) T_B(x -b/2,Y)
\end{equation}
where $\sigma$ is the total inelastic cross-section for nucleon-nucleon pairs at this collision energy, $T_{A,B}$ is the thickness function for a nucleus with A nucleons, and $b$ is the impact parameter.

In the wounded nucleon picture, pairs of interacting nucleons are removed via a combinatorial factor leading to
\begin{equation}
n_{WN} (x,y,b) = T_A(x+b/2,y) \left[ 1 - \left( 1- \frac{\sigma T_B(x-b/2,y)}{B} \right)^B  \right] + {\rm sym}.
\end{equation}
where the symmetry is between the nuclei A and B.
At lower (SPS) energies, it was found that the number of produced particles scaled with the number of wounded nucleons.
For RHIC collisions, one expects that there may be deviation from this scaling due to the increasing importance of hard processes.
Initial conditions are thus scaled partially to the number of wounded nucleons and partially to the number of binary collisions via
\begin{equation}
\epsilon(x,y,b) = \kappa_0\left[\alpha n_{WN}(x,y,b) + (1-\alpha) n_{BC}(x,y,b)\right],
\end{equation}
where $\alpha$ is then the fraction of the initial condition scaled to the number of wounded nucleons. 

We then take this total number of interactions to be proportional to the initial energy density of the system at some time shortly after crossing.
Both the overall energy scale ($\kappa_0$) and the thermalization time ($\tau_0$) will be considered free parameters of the model, in addition to the fraction of the initial condition scaled to the number of wounded nucleons ($\alpha$).
In this paper we take $\tau_0 = 0.8$ fm/c, $\alpha = 0.85$, and for the chosen value of $\kappa_0$, the central energy density at $b=2.21$ fm is $\approx19.5$ GeV/fm$^3$. The choice was made to have the energy scale with $n$, rather than entropy, because energy density more justifiably scales with the number of independent superimposed sources.  These parameters are chosen arbitrarily for the current model, as their values will be the subject of a future publication.

The optical Glauber prescription sets the shape of the initial energy density in the transverse plane, it does not lend any insight into the longitudinal initial conditions.
For that we turn to the experimental data, which shows little deviation from a simple Gaussian profile in longitudinal rapidity.
Since Gaussians tend to remain roughly Gaussian in a hydrodynamic expansion  \cite{Wong:2008ex}, we choose a Gaussian with a radius to be determined along with the other parameters of the model.
For the results presented here, this radius was chosen to be 1.4 units of spatial rapidity.
This choice of longitudinal source is in stark contrast with the long flat region with steep half-Gaussian used in other models \cite{Schenke:2011bn, Schenke:2010nt, Schenke:2010rr, Hirano:2002hv}.

Furthermore, while the Glauber model has proven useful in describing the initial state of the matter, a competing model (CGC) based on the assumption that the low momentum particles at midrapidity are generated by color fields has also been considered by many modelers \cite{McLerran:1993ni, Kharzeev:2000ph, Kharzeev:2002ei, Dumitru:2007qr}. Generally, CGC initial conditions have larger initial eccentricities and larger gradients in the tails of the transverse energy density, thus requiring somewhat larger shear viscosity to match experimental results \cite{Romatschke:2009im}.

With the energy density set, there remain nine components of the initial condition of the stress energy tensor to be set. In determining the initial condition for the collective velocity, hydrodynamic models have taken the approach that given a system which is newly thermalized, one might expect no transverse collective flow to be present \cite{Schenke:2011bn, Schenke:2010nt, Schenke:2010rr, Heinz:2002un, Romatschke:2009im, Kolb:2003dz}.
However, since some finite evolution of the system occurs after the crossing but before thermalization, it is unlikely that no evolution of the stress energy tensor should occur.
If Bjorken boost invariance is valid, and if the dynamics are driven by a traceless stress energy tensor, one can predict the initial transverse acceleration regardless of the underlying theory \cite{Vredevoogd:2008id}. For small $\tau$ we assume the initial flow is a fraction $\xi$ of this result,
We will utilize the result
\begin{equation}
\frac{T^{0x}}{T^{00}} = \frac{- \partial_x T^{00}}{2 T^{00}} \xi \tau,
\end{equation}
where $\xi$ is introduced as a free parameter of the model.
In this paper we take $\xi = 0.5$, or half of the pre-equilibrium flow predicted.
In the longitudinal direction, we will assume that the matter is not moving relative to the Bjorken expansion.

Finally, the initial values of the shear tensor must be set.
In the Landau frame, the shear tensor is constrained by its orthogonality to the fluid velocity ($u_\mu \pi^{\mu \nu} = 0$),  traceless-ness ($g_{\mu \nu} \pi^{\mu \nu}=0$), and symmetry ($\pi^{\mu \nu} = \pi^{\nu \mu}$).
This leaves five components to be determined.
In recent history, other models have either set all the components to zero initially \cite{Schenke:2011bn, Schenke:2010rr} 
 or set all the components to their Navier-Stokes values \cite{Song:2007ux}.
While using the Navier-Stokes values have been more prevalent, this approach leads to increasing values of $[\pi^{xx}/P] (x)$ for large $x$, because $\eta/P \propto T^{-1}$ at low temperature.
Even for moderately small viscosities ($\eta/s \approx 0.16$) and relevant temperatures ($T \approx$ 100 MeV), one finds values of $|\pi^{zz}|$ that approach not just the pressure but the energy density.
If $\epsilon + p + \pi^{zz} = 0$ the conservation of longitudinal momentum approaches singularity and the longitudinal fluid velocity grows without bound. For initial conditions that reproduce experimental results, these problems tend to occur well outside of the freezeout surface. Also, since the relaxation times are very large when the viscous corrections are large, the evolution will tend not to approach such singularities if the initial conditions are not close to them. Because relaxation times tend to be short inside the freezeout surface, the system loses its memory of the initial values of the shear tensor within $\sim 1.0$ fm/c of the onset of hydrodynamics. 
For this study, the initial value of $\pi^{xx}/P$ is chosen independent of position, leaving the effects of varying the parameter as a topic for future study.

\section{Algorithm Description}

As discussed above, the model is designed to explore predictions of (3+1)-dimensional viscous hydrodynamics using averaged initial conditions.
Since this avoids encountering the small-scale fluctuations of event-by-event hydrodynamics, the code does not invoke any aggressive scheme for solving the conservation equations in the presence of singularities or shocks.
Instead, the ten equations of motion (Appx. B) are solved simultaneously using the second order Runge-Kutta (predictor-corrector) method on a fixed mesh in the expanding, Bjorken coordinates.
Spatial derivatives of the integration variables are determined by second order central finite differences, and are corrected for the collective flow and the Bjorken expansion (Appx. A).

The set of variables to be integrated are: the energy density, the fluid velocity relative to the Bjorken expansion, and the five components of the shear tensor in the matter frame, $a_i$.
One of the advantages of choosing matter frame variables is that restrictions to the maximum deviation from equilibrium are more easily applied, though such restrictions are not used in this paper.
The algorithm verifies the conservation the integral of $T^{00}$ in the mesh frame:
\begin{eqnarray}
0 &=& \displaystyle\int d^4x d_\mu T^{\mu 0} = \displaystyle\int d\tau d\eta dx dy \left[ \partial_\tau T^{00} + \partial_i T^{0i} + \frac{T^{00} + T^{zz}}{\tau} \right] \\
C &=&  \displaystyle\int  T^{00} \tau d\eta dx dy   + \displaystyle\int d\tau \left[ \displaystyle\oint T^{0i} ds^i  + \displaystyle\int \frac{T^{zz}}{\tau} \tau d\eta dx dy \right]  \label{Eq:consT}
\end{eqnarray}
where the second term in Eq. \ref{Eq:consT}  is the outgoing flux of $T^{0 i}$ through a surface element $ds^i$, the second term is the work done by the expansion of the mesh, and $C$ is a constant.
This quantity is conserved at the level of a tenth of percent in the three-dimensional viscous case for the cell densities used in this paper, which are $dx = dy = d\eta_s = 0.15$ (fm) and $d\tau = 0.05$ fm/c. 
A simple example of further investigation into the veracity of the code is provided in Appx. C.

\section{Equation of State}

Significant progress on the transition region equation of state has been made since the early days of hydrodynamics \cite{Borsanyi:2010cj, Bazavov:2009zn, Huovinen:2009yb}.
Studies of lattice predictions have shown that adjusting for extrapolation to physical masses leads to a less dramatic phase transition at zero baryon chemical potential than originally assumed.
This leads to an equation of state that is stiffer (higher speed of sound) through the transition region.
The results presented use the three-flavor equation of state from the Wuppertal group  \cite{Borsanyi:2010cj}, but  fits to MILC collaboration \cite{Huovinen:2009yb} data have also been explored.

Of vital importance to the accuracy of transition to the hadronic cascade will be the matching of the energy density at the freeze-out temperature.
While the above equation-of-state fits roughly reproduce data in the hadronic region, the energy density can disagree on the order of 10\% with computations from the hadronic cascade itself.
To ensure that this does not occur, we use the equation of state extracted directly from the hadron gas below the freeze-out temperature ($T_f$), the lattice data above some temperature ($T_l$), and in between will merge the two. We merge continuously in the unitless entropy density $\sigma(T) = s/T^3$ via the weighted linear function, $\sigma = w \sigma_l + (1-w)\sigma_h$ with weight
\begin{equation}
w(T) =  \frac{1}{2} \left[ \tanh{ \left( \tan{  \left[ \frac{\pi}{2} \left(  2\frac{T_l - T}{T_l - T_f} -1  \right) \right]} \right)}  + 1 \right].
\end{equation}
which ensures the continuity and smoothness of $\sigma(T)$ and therefore a continuous speed of sound squared via
\begin{equation}
c_s^2 = \frac{ 1}{ 3 +\frac{T}{\sigma} \frac{ d \sigma}{dT}}.
\end{equation}

\begin{figure}[fig:eos]
\centerline{\includegraphics[width=0.6 \textwidth]{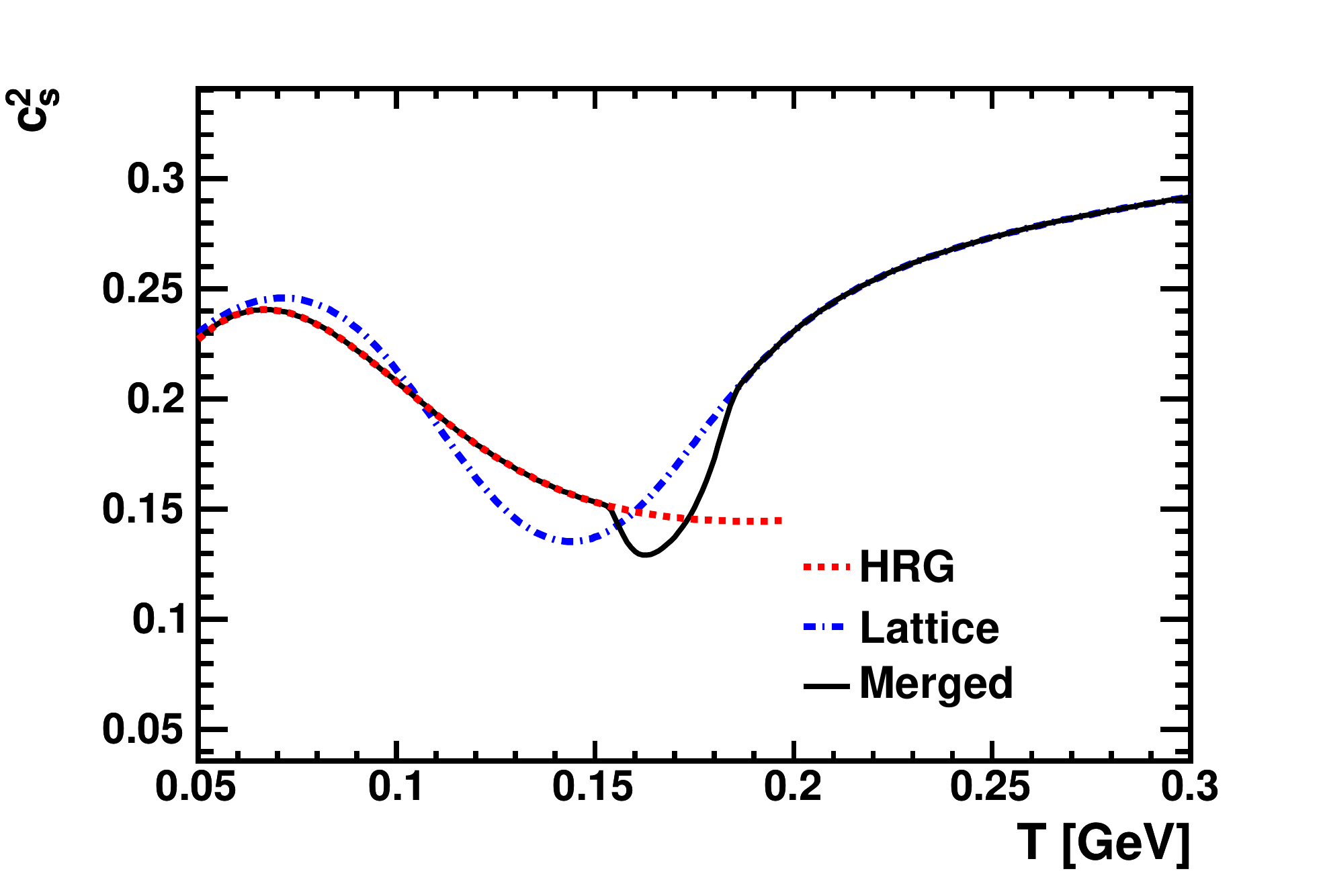}}
\caption{ (color online)
The squared speed of sound as a function of the temperature from the hadron resonance gas (red dotted), a fit to lattice data  \cite{Borsanyi:2010cj} (blue dashed), and the merging of the two (black solid).
The fit to the lattice data predicts entropy production at a lower temperature than observed in the hadron gas as seen in the lower speed of sound.
Matching the entropy at $T_l$ then requires a soft region between $T_h$ and $T_l$.
}
\label{fig:eos}
\end{figure}
We take $T_l = 190$ GeV and $T_f = 150$ GeV.  
This choice of parameters leads to significant entropy production at temperatures immediately above the freeze-out surface, which is observed as a soft region in the speed of sound.

Once the entropy density has been determined at all relevant temperatures, $p(T)$ is evaluated via the integral
\begin{equation}
p(T) = p(T_f) +  \int_{T_f}^T s(T') dT'
\end{equation}
This procedure was employed to generate $\epsilon$, $p$, and $s$ at a spacing of 1 MeV in temperature.
The equation of state was then smoothed via a traditional cubic spline \cite{Press:1992} to ensure consistency and smoothness in the interpolation.

The only remaining thermodynamic variables to be determined are related to the viscosity of the matter.
It is generally thought that away from the cross-over temperature the ratio of shear viscosity to entropy density should be larger, although most models have fixed this ratio for all temperatures. 
However, viscous suppression of anisotropic flow at RHIC seems mostly to be driven by the value of the shear viscosity at the freeze-surface surface  \cite{Niemi:2011ix}
although this may not be true at LHC energies \cite{Niemi:2011nd}.
In some cases, the low temperature regions of this model were found to be unstable with this choice and so at very low temperatures the shear viscosity was chosen to continuously vary with the energy density below some temperature well outside of the freeze-out surface ($\approx 120$ MeV).  
The model was found to be insensitive to the choice of this temperature as long as the difference between it and the freeze-out temperature was more than 20 MeV.

Finally, we take the relaxation time to be $\tau_\eta = \frac{ 3 \eta}{Ts}$.
This result can derived from linear response theory under the assumptions that the field loses correlation and that the entropy penalty for small viscous corrections is quadratic  \cite{Pratt:2007gj}.

\section{Results}

We consider three centrality classes corresponding to three impact parameters.
For the 0-5\% centrality class we use $b=2.21$ fm, $b=5.70$ fm for 10-20\%, and $b=7.23$ for 20-30\%, which are the values determined by the STAR collaboration using a Monte Carlo Glauber model \cite{:2008ez}.
The energy scale, $\kappa_0$, fixes the maximum energy density in the most central collisions (b=2.21 fm) to be 19.5 GeV/fm$^3$ at $\tau_0 = 0.8$ fm/c.
The shear viscosity to entropy density ratio is taken to be 0.16, the initial value of the longitudinal part of the shear tensor $\tilde{\pi}^{zz}= -p/2$, and the initial transverse flow parameter is taken as $\xi = 0.5$.  

\begin{figure}
\centerline{\includegraphics[width=0.6 \textwidth]{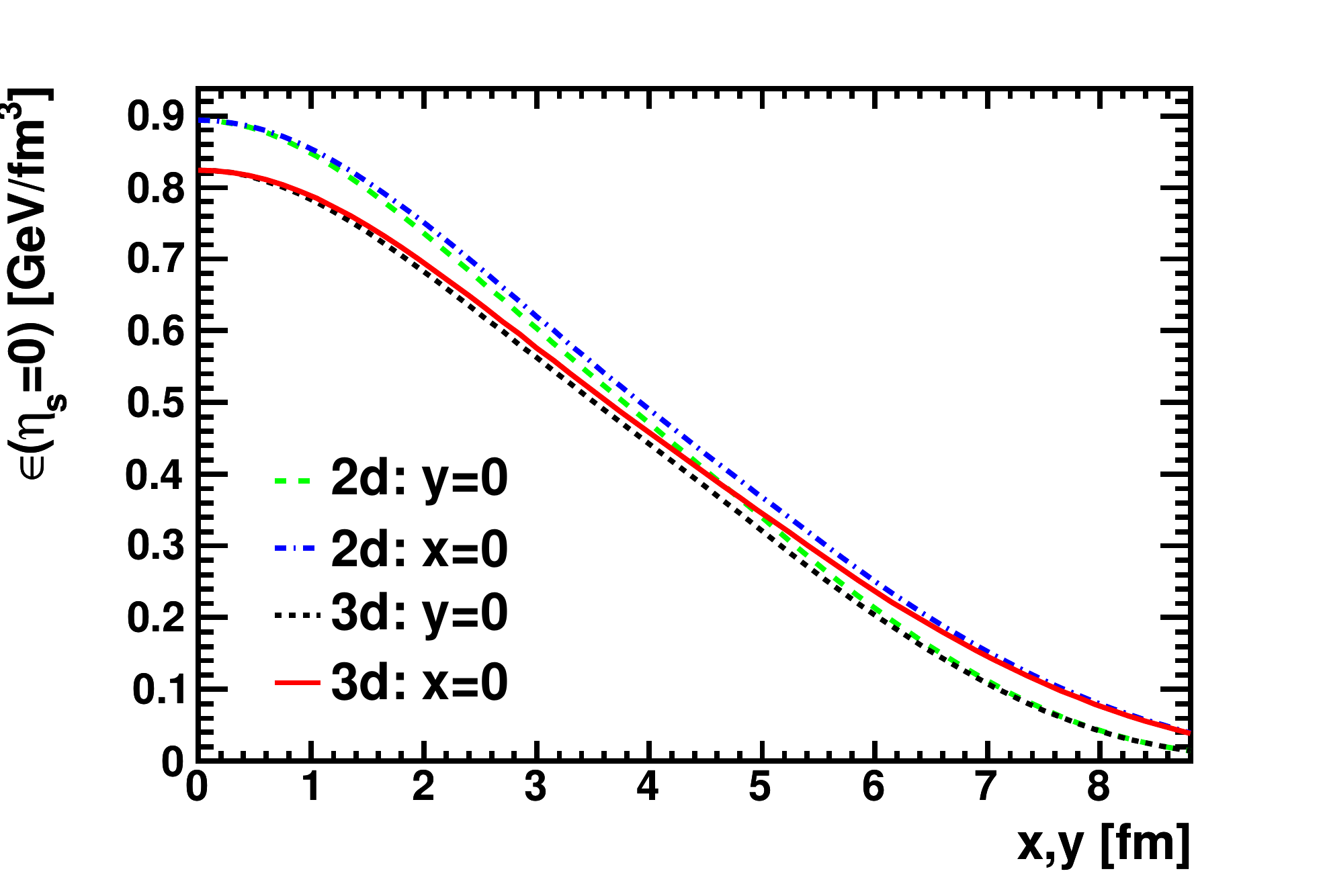}}
\caption{ (color online) 
Snapshot of the energy density at $\eta_s = 0$ and $\tau = 5.52$ fm/c along the $x$ and $y$ axes with and without Bjorken invariance for b = 5.70 fm.
The central energy density falls more rapidly without Bjorken invariance, but the effect on the evolution diminishes in the periphery.
}
\label{fig:dens}
\end{figure}

We begin by demonstrating the effects of the finite longitudinal extent on the hydrodynamic evolution.
Figure \ref{fig:dens} shows the energy density at a fixed time along the transverse axes at mid-rapidity.
The energy density near the center of the fireball falls somewhat faster in the full three-dimensional model.  This was expected given that the additional longitudinal acceleration, which is neglected in a boost-invariant expansion, which leads to a slightly more rapid expansion and lower energy densities.
However, the evolution in the periphery has changed surprisingly little.

\begin{figure}
\centerline{\includegraphics[width=0.6 \textwidth]{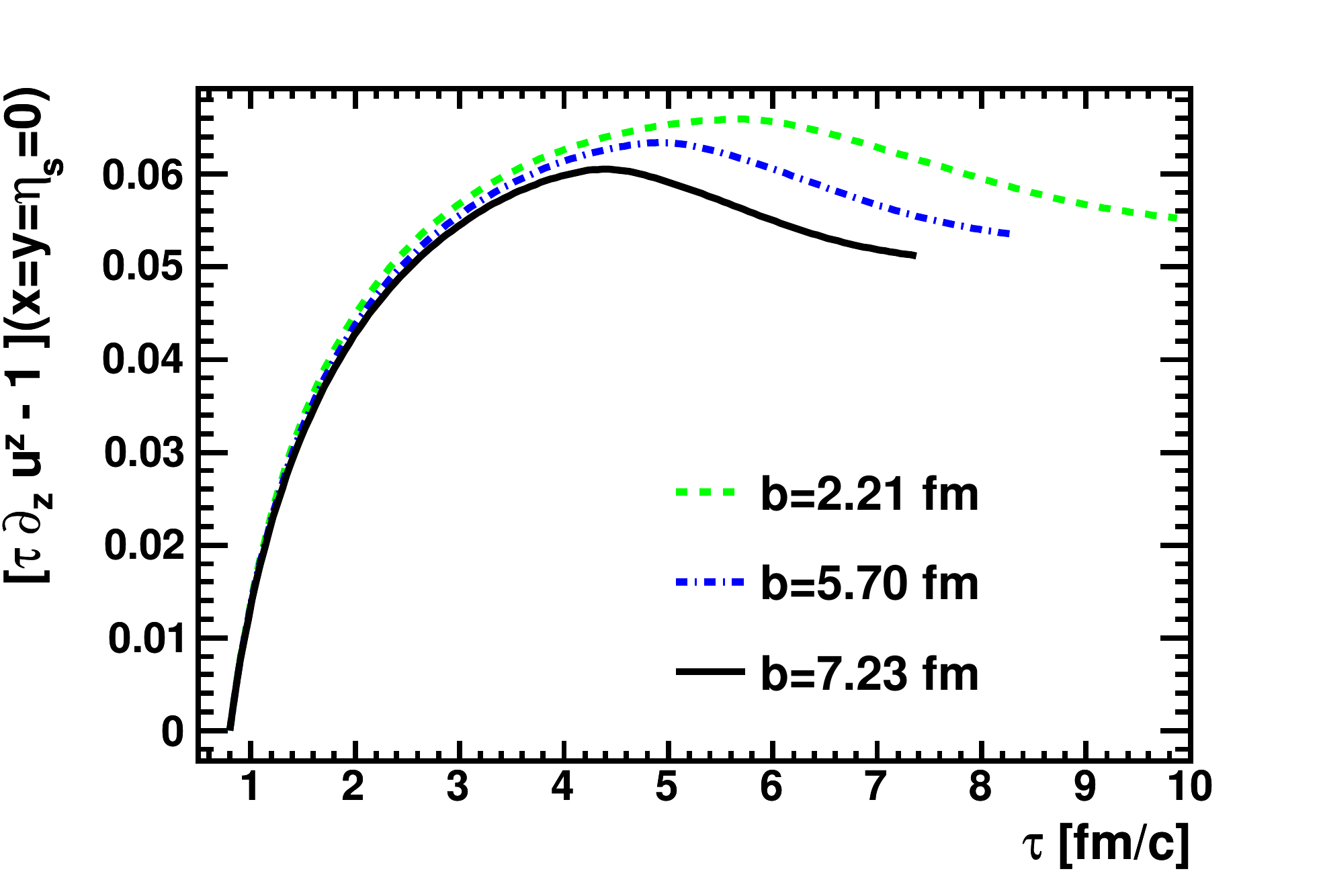}}
\caption{ (color online)
The longitudinal velocity gradient at the origin equals $1/\tau$ for a pure Bjorken expansion, so the ratio above is explicitly zero in the two-dimensional treatments.
The velocity gradient varies from the Bjorken assumption in the three dimensional model on the order of 5\%.
}
\label{fig:dzz}
\end{figure}

Figure \ref{fig:dzz} quantifies the extent to which the velocity gradient at the origin falls faster in the full three-dimensional model. Whereas the velocity gradient is $\partial_z u^z=1/\tau$ in the boost-invariant picture, longitudinal acceleration increases the gradient. The increase from the Bjorken value of the longitudinal velocity shows little dependence on the impact parameter and is at the five percent level. This suggests that femtoscopic estimates of lifetime based on the boost-invariant assumption are a few percent low. This discrepancy should be somewhat larger for smaller viscosities, because an increase in transverse pressure should increase the longitudinal acceleration.

\begin{figure}
\centerline{\includegraphics[width=0.6 \textwidth]{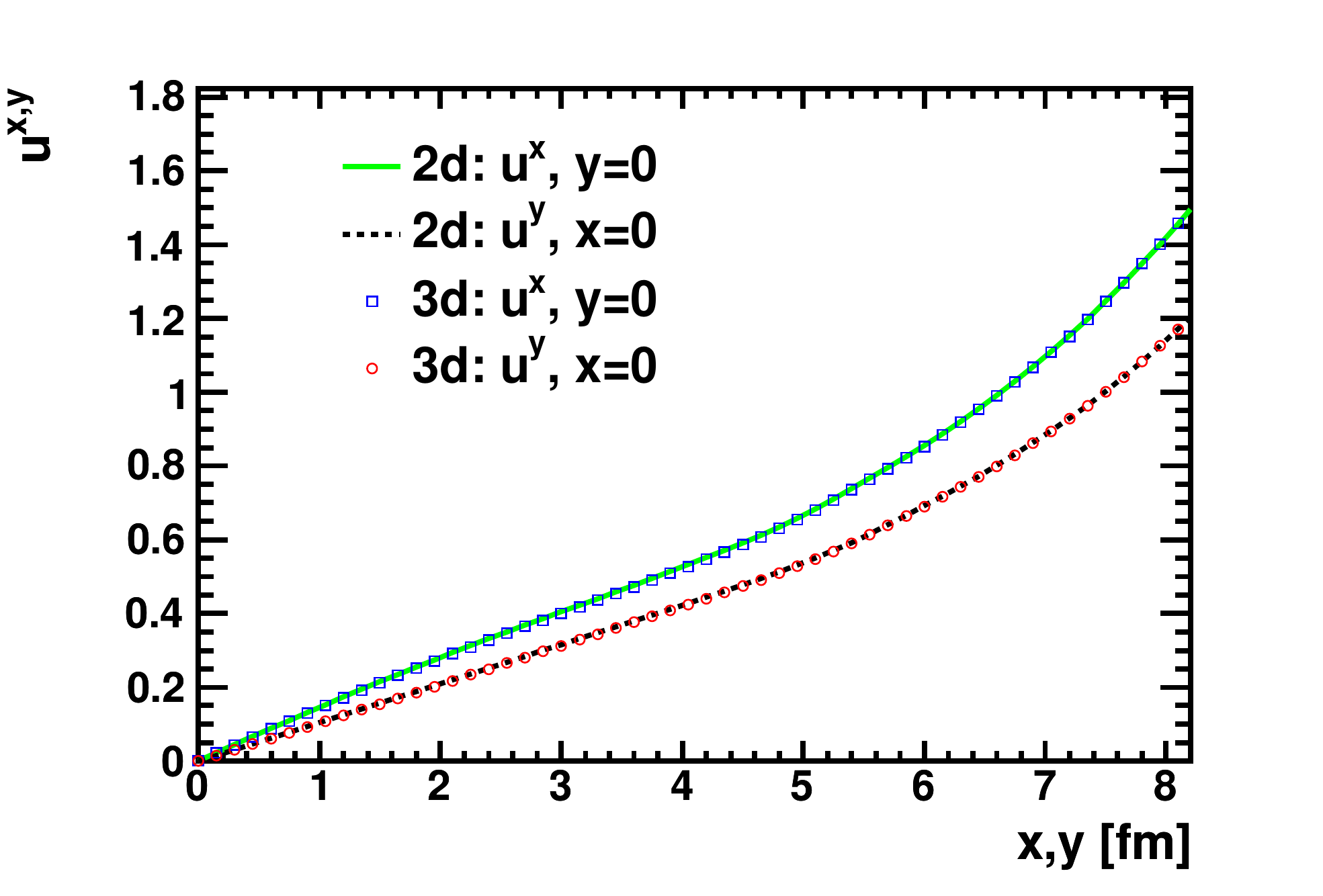}}
\caption{ (color online) 
Snapshot of the collective at $\eta_s = 0$ and $\tau = 5.52$ fm/c along the $x$ and $y$ axes with and without Bjorken invariance for b = 5.70 fm.
The collective velocity at $\eta_s=0$ is found to be unaffected by the non-trivial longitudinal expansion.
}
\label{fig:uVec}
\end{figure}

Figure \ref{fig:dzz} suggests that the change in the energy density profile at $\eta_s = 0$ is due to the presence non-trivial longitudinal expansion,
and not due to changes in the transverse flow.
In Figure \ref{fig:uVec}, we turn our attention to this transverse flow under the same conditions as Figure \ref{fig:dens}.
Whereas the energy density differed by up to 5\% at midrapidity, we find that the collective velocity differs by a maximum of 1\% for the same initial conditions.
This implies that any significant modification to the elliptic flow at midrapidity will not come from changes in the dynamics at $\eta_s =0$,
though we cannot address changes to the initial conditions require to fit multiplicities or the effects of thermal smearing in this study.

\begin{figure}[fig:fos]
\centerline{\includegraphics[width=0.6 \textwidth]{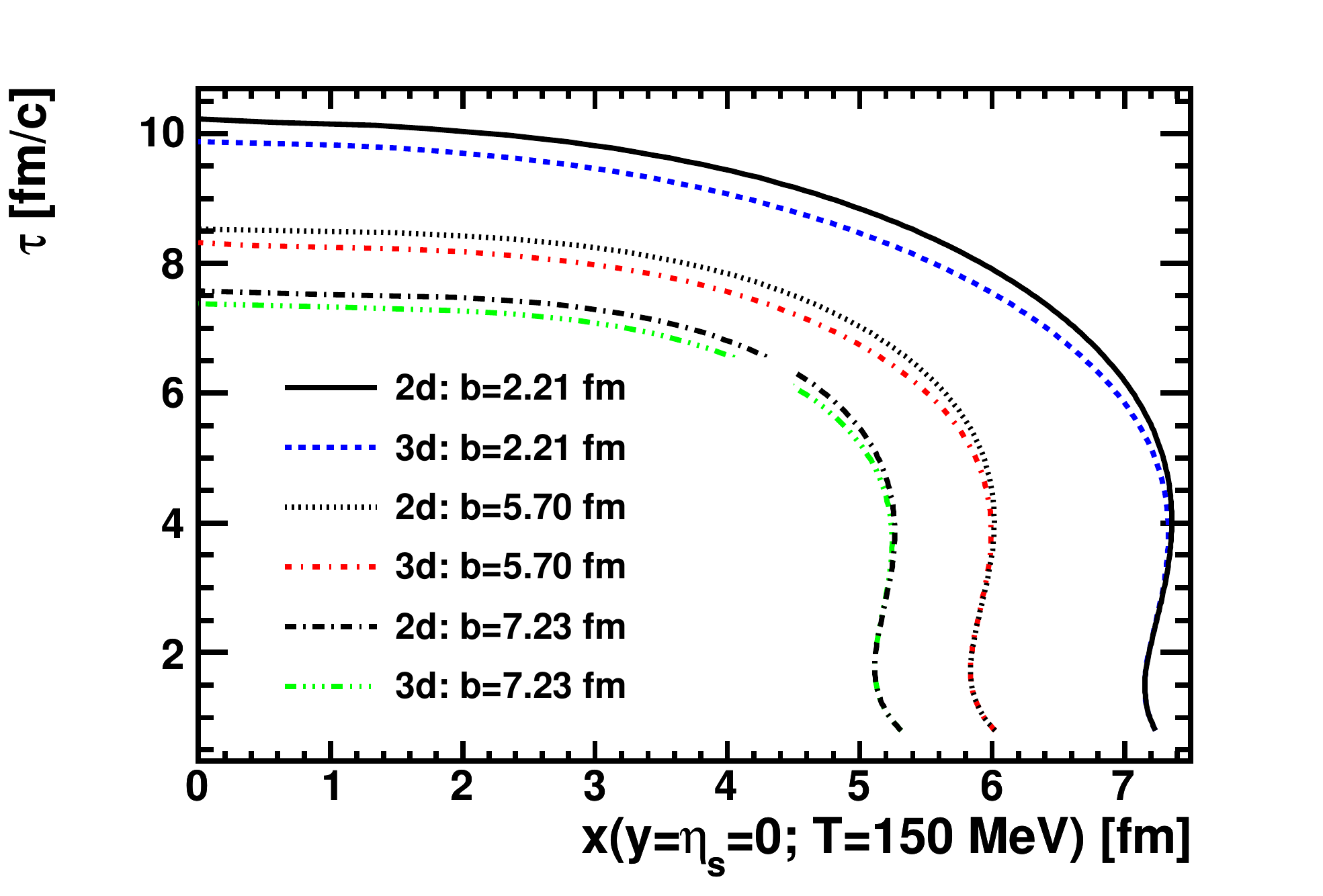}}
\caption{ (color online)
The position of the freeze-out surface along the x-axis at mid-rapidity for three impact parameters. 
Incorporating the full three-dimensional motion reduces the lifetime of the source by about 5\%.
However, since much of the emission comes from the sudden, final collapse of the source the effect on emission may be modest.
}
\label{fig:fos}
\end{figure}

Figure \ref{fig:fos} shows the difference in the location of the freeze-out surface along the x-axis at mid-rapidity. Due to the increased longitudinal velocity gradient in the three-dimensional calculation, the system breaks up slightly faster. All of these changes suggest only modest changes to the mid-rapidity observables predicted by boost-invariant viscous hydrodynamics.

\section{Conclusion and Outlook}

We have presented a new simulation of (3+1) dimensional Israel-Stewart viscous hydrodynamics for relativistic heavy ion collisions geared toward predictions using average initial conditions. 
The two-dimensional version of this model was found to be highly consistent with two-dimensional calculations from other groups. 
The algorithm conserves $T^{00}$ at a high level, 
 and in the limit of zero viscosity, entropy was conserved at better than the 0.1\% level.

Investigations of the effect of the non-trivial longitudinal expansion show the strongest effect at the level of 5\%, 
as the system expands and disintegrates modestly faster than in corresponding two-dimensional calculations.
The effects of the non-trivial expansion are stronger near the center of the fireball than in the periphery.
Combined with very small changes to the transverse velocities, this causes the motion of the freeze-out surface to remain almost unchanged until later times when it falls more rapidly.
This suggests that changes to important midrapidity observables like elliptic flow will not be larger than a few percent.


These results suggest that one can ignore the effects of the full three-dimensional expansion if one is satisfied with predictions at the 5\% level. 
Given that many approximations in the model are only trustworthy at the 5\% level, these corrections can be considered modest, but cannot be classified as fully negligible. 
If one were to repeat the comparison at lower energies, the effects would certainly be greater, while at higher energies such as at the LHC, boost invariance should be significantly more justified.

We intend to implement a freeze-out algorithm in the near future. 
Currently, the interface between the cascade and the hydrodynamic model only functions in the two-dimensional version of the code. 
Once the three-dimensional interface is finished, comparison with data will be performed with observables at and away from mid-rapidity.

\begin{acknowledgments}
J.V. would like to recognize Steffen Bass, Pasi Huovinen, Gabriel Denicol, Harri Niemi, and Bj\"oern Schenke for sharing their insight during the course of this project, 
and to HIC for FAIR (20130403) for funding a 2010 visit when much of this work was discussed.
Support was provided by the U.S. Department of Energy, Grant No. DE-FG02-03ER41259. 

\end{acknowledgments}

\appendix

\section{Evaluation of Local Derivatives} 
For our set of integration variables, we choose a mix of quantities measured in the frame moving with the energy density and quantities measured in a frame undergoing a fixed expansion in spatial rapidity, which we refer to as the mesh frame.
Beginning from the equations of motion as viewed in the matter frame, we must boost terms into the mesh frame.
For instance, to track the velocity as observed in the mesh frame it is prudent to evaluate equations of motion in terms of spatial derivatives of the collective velocity in the mesh frame.
In addition, one needs to write the equations of motion in terms of derivatives evaluated on a fixed grid of the mesh frame proper time.

We introduce the general form of a boost from a frame that observes some velocity $u^\mu$ to one that observes any other velocity $n^\mu$.
In terms of these velocities and the metric tensor ($g^{\mu \nu}$), the boost is:
\begin{equation}
\Lambda^{\mu \nu} (u \rightarrow n) = g^{\mu \nu} + 2 n^\mu u^\nu - \frac{ (u^\mu + n^\mu) ( u^\nu + n^\nu)}{1+u \cdot n}
\end{equation}
which fulfills the requirement that $\Lambda^{\mu \nu} u_\nu = n^\mu$.

This boost formula can be used to evaluate local derivatives of local velocities.  
We take $n$ to be the velocity as viewed in the matter frame, which is by definition $n^\mu = (1,0,0,0)$, and $u$ to be the velocity as viewed in the mesh frame.
Likewise, we define the local derivative as $\tilde{\partial}_\mu = \Lambda_\mu^\nu \partial_\nu$.
Then by one method, we can evaluate local derivatives of local velocities by simply evaluating:
\begin{eqnarray}
\tilde{\partial_\alpha} n^\beta & = & \Lambda^\beta_\nu \tilde{\partial}_\alpha \text{\emph{u}}^\nu  \nonumber \\
&=& \left[ g^\beta_\nu + 2 n^\beta u_\nu - \frac{(u^\beta + n^\beta)(u_\nu + n_\nu)}{1+\gamma}  \right] \tilde{\partial}_\alpha u^\nu \\
&=& \tilde{\partial}_\alpha u^\beta - \frac{u^\beta + n^\beta}{1+\gamma} \tilde{\partial}_\alpha \gamma 
\end{eqnarray}
which can be used to evaluate both spatial and temporal derivatives of the collective velocity.
For instance, local spatial derivatives can be expressed
\begin{eqnarray}
\tilde{ \partial}_i n^j & = & \partial_i u^j + u^i   \dot{u}^j + \frac{u^i u^k}{1+\gamma} \partial_k u^j \nonumber \\
&&  - \left( \frac{  u^j u^k}{\gamma ( 1+ \gamma) } \right) \left[ \partial_i u^k + u^i \dot{u}^k + \frac{u^i u^m}{1+\gamma} \partial_m u^k  \right]
\end{eqnarray}
where latin indices run over the spatial coordinates and repeated indices are summed over.
The structure of our code is such that the coefficients associated with time derivatives of tabulated variables are separated from spatial derivatives, leading to the notation
\begin{equation}
\tilde{ \partial}_i n^j = (\tilde{ \partial}_i n^j )' + u^i   \dot{u}^j - \left( \frac{  u^j u^k}{\gamma ( 1+ \gamma) } \right) u^i \dot{u}^k
\end{equation}
where the prime indicates the removal of those terms proportional to time derivatives of tabulated variables.

Also among our variables are terms defined in the frame of the collective motion.
The above construction cannot be used to calculate derivatives of these variables.
Instead, we consider a small boost in the neighborhood of the boost from the mesh frame to the fluid frame.
In our notation, this will be referred to as $\delta \Lambda(u - \delta u,n)$ where the choice of sign allows us to treat the differential element as naturally covariant (where the differential operator is normally naturally contravariant).
Reducing to only terms linear in the deviation of the fluid velocity:
\begin{eqnarray}
\delta \Lambda^{\mu \nu} (u - \delta u , n) & = & g^{\mu \nu} + 2 n^\mu (u^\nu - \delta u^\nu)  - \frac{ (u^\mu - \delta u^\mu + n^\mu) (u^\nu - \delta u^\nu + n^\mu)}{1 + \gamma - \delta \gamma} - \Lambda(u,n) \\
&=& -2 n^\mu \delta u^\nu + \frac{ (u^\mu + n^\mu ) \delta u^\nu}{1 + \gamma} + \frac{ \delta u^\mu  (u^\nu + n^\nu )}{1 + \gamma}  - \frac{\delta \gamma}{(1+\gamma)^2} (u^\mu + n^\mu) ( u^\nu + n^\nu) \nonumber
\end{eqnarray}
This method confirms the above results for local derivatives of the collective velocity by noting that 
\begin{equation}
\delta \Lambda^\mu_\alpha u^\alpha = \delta u^\mu - \frac{u^\mu + n^\mu}{1+\gamma} \delta \gamma
\end{equation}
where we have frequently used the orthogonality relationship $u^\mu \delta u_\mu = 0$.

Now, since we wish to obtain corrections to derivatives of a tensor in the matter frame, a boost back to the frame of the matter is required.
The boost to be applied to the matter frame tensor is then:
\begin{equation}
\delta \Lambda^{\mu \alpha}  \Lambda_{\alpha}^\nu
= 2 (n^\mu \delta u^\nu - \delta u^\mu u^\nu) 
+ \frac{2 \delta \gamma}{1+\gamma} \left[u^\mu n^\nu - n^\mu u^\nu \right] 
+ \frac{1}{1+\gamma} \left[ \delta u^\mu (u^\nu + n^\nu) - (u^\mu +n^\mu) \delta u^\nu \right] 
\end{equation}
which can be applied symmetrically to the local shear tensor:
\begin{eqnarray}
\delta \pi^{\mu \nu} &=&  \delta \Lambda^{\mu \alpha}  \Lambda_{\alpha \beta} \pi^{\beta \nu}  + \delta \Lambda^{\alpha \nu} \Lambda_{\alpha \beta} \pi^{\mu \beta} \nonumber \\
\delta \Lambda^{\mu \alpha}  \Lambda_{\alpha \beta} \pi^{\beta \nu} & = &  2  n^\mu \delta u_\alpha \pi^{\alpha \nu} + \frac{1}{1+\gamma} \left[ \delta u^\mu u_\alpha \pi^{\alpha \nu} - (u^\mu + n^\mu) \delta u_\alpha \pi^{\alpha \nu} \right] - \frac{2 \delta \gamma}{1+\gamma} n^\mu u_\alpha \pi^{\alpha \nu}
\end{eqnarray}
where we have used the choice of frame occasionally ($n_\mu \pi^{\mu \nu} = 0$).

If both indices on the shear tensor are time-like, $\delta \pi^{0 0} = 0$. 
For one time-like index and one space-like index ($i$), 
\begin{eqnarray}
\delta \pi^{0 i} &=& \delta \Lambda^{0 \alpha} \Lambda_{\alpha \beta} \pi^{\beta i} + \delta \Lambda^{i \alpha} \Lambda_{\alpha \beta} \pi^{\beta 0} \nonumber \\
&=& \delta u_\alpha \pi^{\alpha i} - \frac{\delta \gamma}{1+\gamma} u_\alpha \pi^{\alpha i} = - \pi^i_\alpha \delta n^\alpha 
\end{eqnarray}
where the final equivalence can be seen as a consistency check.
Finally, for two spatial indices,
\begin{equation}
\delta \pi^{ij} = \frac{1}{1+\gamma} \left[ \delta u^i u_\alpha \pi^{\alpha j} + \delta u^j u_\alpha \pi^{\alpha i} - u^i \delta u_\alpha \pi^{\alpha j} - u^j \delta u_\alpha \pi^{\alpha i} \right] 
\end{equation}

As an example, consider the simplified Israel-Stewart equation
\begin{equation}
\tau_\pi \tilde{\partial}_0 ( \pi^{\mu \nu}) + \pi^{\mu \nu}  = \pi^{\mu \nu}_{(NS)}  
\end{equation}
After boosting the time derivative from the frame of the matter to the mesh frame via $ \tilde{\partial}_0 = \gamma \partial_0 + u^i \partial_i $,
there remains a co-moving time derivative of a spatial component of the local shear tensor ($ \gamma \partial_0 \pi^{ij} $).
Simply integrating forward with respect to time would produce errors due to the changing reference frame in which $\pi$ is defined.
Therefore, one amends this via
\begin{equation}
\partial_0 \pi^{ij} \rightarrow \dot{\pi}^{ij} + \frac{1}{1+\gamma} \left[  u^i \dot{u}^k \pi^{kj} + \pi^{ik} \dot{u}^k u^j - \dot{u}^i u^k \pi^{kj} - \pi^{ik} u^k \dot{u}^j   \right]
\end{equation}



Now, our calculation is based on a fixed mesh in the expanding coordinate system $(\tau, x, y , \eta)$.
Given this coordinate system, it only makes sense to choose our variables to be defined in the fixed frame of this coordinate system (the mesh frame).
So when discussing the collective velocity $u$, we refer not to the collective velocity observed in the laboratory frame - the center of mass frame for the collision - but rather to the collective velocity observed by a comover with the longitudinal expansion.
However,  our equations of motion will be derived in the frame of the matter - a third, potentially unique Lorentz frame.
In light of this, our approach will again be to determine the small correction to be applied to derivatives in the mesh frame and boost this correction back into the matter frame.

We begin by considering the velocity as observed in the mesh frame $u^\mu = \{\gamma, u^x, u^y, u^z \}$.
If we were to view this vector in the laboratory frame ($u_l^\mu$), we would observe instead 
\begin{equation}
u_l^\mu = \{\gamma \cosh{\eta} + u^z \sinh{\eta}, u^x, u^y, u^z \cosh{\eta} + \gamma \sinh{\eta}  \}
\end{equation}
So, viewed from a frame at $\eta = \delta \eta $, $ \gamma = \gamma + u^z \delta \eta $ and $ u^z = u^z + \gamma \delta \eta $.
This means that if between mesh cells at slightly differing spatial rapidity one observes that $u^z(0) = u^z(\delta \eta)$, the difference in frame dictates that $\partial_\eta u^z(0) = \gamma$.

Now, any particular mesh frame may accurately view itself as being at $\eta = 0$, and its relationships to its neighbors at slightly differing spatial rapidity are preserved.
So if one considered a small boost $\delta \Lambda^{\mu \nu} = n^\mu \eta^\nu - \eta^\mu n^\nu$ where $n^\mu = \{ 1, 0, 0, 0\} $ and $\eta^\mu = \{ 0, 0, 0, -\delta \eta \}$,
such a boost would adjust the observed velocity as required:
\begin{equation}
\delta \Lambda^\mu_\alpha u^\alpha  = (n^\mu \eta_\alpha - \eta^\mu n_\alpha) u^\alpha = n^\mu (u \cdot \eta) - \gamma \eta^\mu = \delta^{\mu 0} u^z \delta \eta + \delta^{\mu \eta} \gamma \delta \eta
\end{equation}

Boosting this into the frame of the matter,
\begin{eqnarray}
\Lambda^{\mu \alpha}(u,n) \Lambda^{\nu \beta} (u,n) \delta \Lambda_{\alpha \beta} &=& \Lambda^{\mu \alpha} \Lambda^{\nu \beta} (n_\alpha \eta_\beta - \eta_\alpha n_\beta) \nonumber \\
&=& (2 \gamma n^\mu - u^\mu ) ( \eta^\nu + 2 (u \cdot \eta) n^\nu - \frac{u \cdot \eta}{1+\gamma} [u^\nu + n^\nu] ) \nonumber \\
&& - (\eta^\mu + 2 (u \cdot \eta) n^\mu - \frac{u \cdot \eta}{1+\gamma} [ u^\mu + n^\mu] ) (2 \gamma n^\nu - u^\nu)
\end{eqnarray}

Applying this to the matter frame velocity, we obtain several results
\begin{eqnarray}
(\Lambda \Lambda \delta \Lambda)^0_\alpha n^\alpha &=& \gamma (u \cdot \eta) - \gamma (u \cdot \eta) = 0 \\
(\Lambda \Lambda \delta \Lambda)^x_\alpha n^\alpha &=& - u^x (u \cdot \eta) + \frac{\gamma}{1+\gamma} u^x (u \cdot \eta) = \frac{-u^x}{1+\gamma} u^z \delta \eta \\
(\Lambda \Lambda \delta \Lambda)^z_\alpha n^\alpha &=& - u^z (u \cdot \eta) - \gamma \left[ \eta^z - \frac{u^z}{1+\gamma} (u \cdot \eta) \right] = \left[ \gamma - \frac{(u^z)^2}{1+\gamma} \right] \delta \eta 
\end{eqnarray}
and a few results for the shear tensor
\begin{eqnarray}
(\Lambda \Lambda \delta \Lambda)^x_\alpha \pi^{\alpha \nu} &=& - u^x \delta \eta \pi^{z \nu} \\
(\Lambda \Lambda \delta \Lambda)^z_\alpha \pi^{\alpha \nu} &=& - (u^z \pi^{z \nu} -  u_\alpha \pi^{\alpha \nu}) \delta \eta = (u^x \pi^{x \nu} + u^y \pi^{y \nu} ) \delta \eta 
\end{eqnarray}
which will suffice as linear corrections to the shear tensor are always of these forms.


\section{Equations of Motion}

Since the equations of motion here are for a set of variables not previously used, we include a sketch of the derivation of the equations of motion.
In general, we will not include terms relating to the longitudinal expansion but will include time derivatives of the collective velocity related to the motion of the matter frame.

In the frame of the matter, the conservation equations take the form
\begin{eqnarray}
\tilde{\partial_0} \epsilon + (\epsilon+P) ~\tilde{\partial_i} n^i + \pi^{ij} ~ \tilde{\partial}_i n^j  &=& 0\\
(\epsilon + P) ~ \tilde{\partial_0} n^i + \pi^{ij} ~ \tilde{\partial_0} n^j + \tilde{\partial}_i P + \tilde{\partial_j} \pi^{ij} &=& 0  
\end{eqnarray}
Excluded here are any terms proportional to $n^i$, for instance $n^j \tilde{\partial}_i \pi^{ij}$, which are explicitly zero in this frame.

Separating out the time derivatives, but leaving alone $(\tilde{\partial}_i n^j)'$ as defined in Appendix A for energy conservation:
\begin{eqnarray}
\gamma \dot{\epsilon} + \frac{(\epsilon+P) u^i}{\gamma} \dot{u}^i + (u^i \pi^{ij})  \dot{u}^j - \frac{ (u^i u^j \pi^{ij}) u^k}{\gamma(1+\gamma)} \dot{u}^k &&\\
+ u^i \partial_i \epsilon + (\epsilon + P)  \partial_i u^i + \pi^{ij} ( \tilde{\partial}_i n^j )' &=& 0  \nonumber
\end{eqnarray}
where the second term of the first line comes from $\dot{\gamma}$ and final two terms of the first line come from $\tilde{\partial}_i n^j$.

Doing the same for momentum conservation:
\begin{eqnarray}
 u^i c_s^2 \dot{\epsilon} + (\epsilon+P) \gamma \dot{u}^i - \frac{ (\epsilon+P) u^i u^j}{1+\gamma} \dot{u}^j + \gamma \pi^{ij} \dot{u}^j && \\
+ \frac{ u^k }{1+\gamma} \left[ u^i \pi^{kj} - 2 \pi^{ik} u^j  \right]  \dot{u}^j+ (\gamma -1) \pi^{ij} \dot{u}^j   + u^j \dot{ \pi }^{ij} && \nonumber \\
+ (\epsilon + P) (\tilde{\partial}_0 n^i)' + \pi^{ij} (\tilde{\partial}_0 n^j)' + \partial_i P + \frac{u^i u^k}{1+\gamma} \partial_k P  + (\tilde{\partial}_j \pi^{ij})' &=& 0 \nonumber
\end{eqnarray}
where several of the terms at the end of the first line come from $\tilde{\partial}_j \pi^{ij} \rightarrow  u^j \partial_0 \pi^{ij}$.

In the frame of the matter, the Israel-Stewart equation becomes a relaxation equation:
\begin{equation}
\tau_\pi \tilde{\partial}_0 \pi^{ij}  + \tau_\pi (4/3) \pi^{ij} \partial_k n^k = \pi^{ij}_{(NS)}  - \pi^{ij} 
\end{equation}
which, for example, for the first projected shear stress element is:
\begin{equation}
\gamma  \partial_0 {a}_1 +u^i \partial_i a_1  + (4/3) a_1 \partial_\mu u^\mu + \frac{a_1}{\tau_\pi} = \frac{ -\eta}{\tau_\pi} (\tilde{\partial}_x n^x - \tilde{\partial}_y n^y)
\end{equation}
Extracting all of the time derivatives, we obtain
\begin{eqnarray}
\gamma \dot{a}_1 &-& \frac{2}{1+\gamma} \left[ \left( u^x \pi^{xj} - u^y \pi^{yj} \right) \dot{u}^j   - u^j \pi^{jx} \dot{u}^x + u^j \pi^{jy} \dot{u}^y   \right]   \\
&+& \frac{\eta}{\tau_\pi} \left( (u^x \dot{u}^x - u^y \dot{u}^y) - (u_x^2 - u_y^2) \frac{u^k \dot{u}^k }{\gamma(1+\gamma)}  \right) + \frac{4 u^j a_1}{3\gamma} \dot{u}^j \nonumber \\
&=& -u^i \partial_i a_1 - (4/3) a_1 \partial_i u^i - \frac{a_1}{\tau_\pi} - \frac{\eta}{\tau_\pi} [ (\tilde{\partial}_x n^x)' - (\tilde{\partial}_y n^y)' ] \nonumber
\end{eqnarray}
where $\partial_i a_1$ still contains corrections due to velocity gradients.


\section{Exponential Test}

\begin{figure}[fig:expU]
\centerline{\includegraphics[width=0.5 \textwidth]{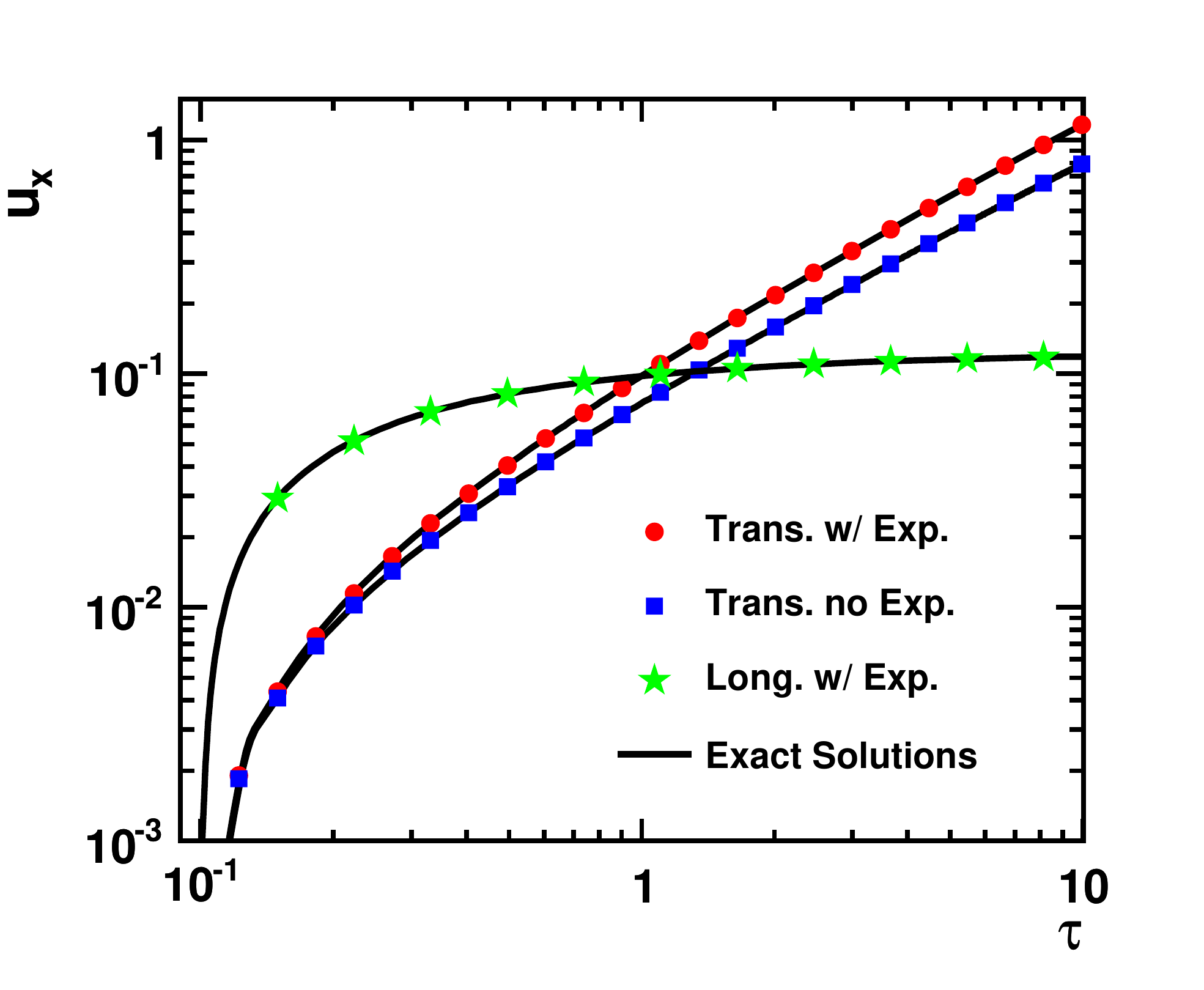}}
\caption{ (color online)
The collective velocity at all points for a exponential energy density profile with no Bjorken expansion (blue squares), a transverse profile with Bjorken expansion (red stars),
and a longitudinal profile with Bjorken expansion (green triangles).
In each case the velocity calculated by the full code agrees with the correct value.
}
\label{fig:expU}
\end{figure}

One of the tests performed to validate the (3+1) dimensional code was the integration of a stationary exponential profile in energy density with length scale $R$:
\begin{equation}
\epsilon = \epsilon_0 \cdot e^{-x/R} , ~~~~~ \vec{u} = 0 \label{expHydro_IC}
\end{equation}
Since the energy gradients are then proportional to their values, the ideal hydrodynamic equations become linear and the collective velocity remains the same at every point.
One may then solve for the transverse collective velocity in the presence of a Bjorken expansion, yielding a simple linear differential equation:
\begin{equation}
\dot{u} = \frac{ u_0 c_s^2}{ u_0^2 - c_s^2 u^2} \left[ \frac{u_0 u}{\tau} + \frac{1}{R(1+c_s^2)} \right] \label{expHydro_UDB}
\end{equation}
where the first term is from the Bjorken expansion and the second is from the exponential gradient.
In the case where there is no Bjorken Expansion, the first term is zero and the term equation has an analytical solution:
\begin{equation}
 \frac{c_s^2}{R(1+c_s^2)}t = \frac{1}{2} \left[ (1+c_s^2) \sinh^{-1}(u) - (c_s^2-1)u\sqrt{u^2+1} \right] \label{expHydro_AnalyticU}
\end {equation}
In the presence of the Bjorken expansion the equation is not separable and no analytical solution was pursued.
The same computation for a longitudinal exponential profile yields the same differential equation with the assignment $R \rightarrow \sigma_\eta \tau$ in the solution.
The solutions for the these three cases are presented in Fig. \ref{fig:expU}.

\begin{figure}[fig:expE]
\centerline{\includegraphics[width=0.5 \textwidth]{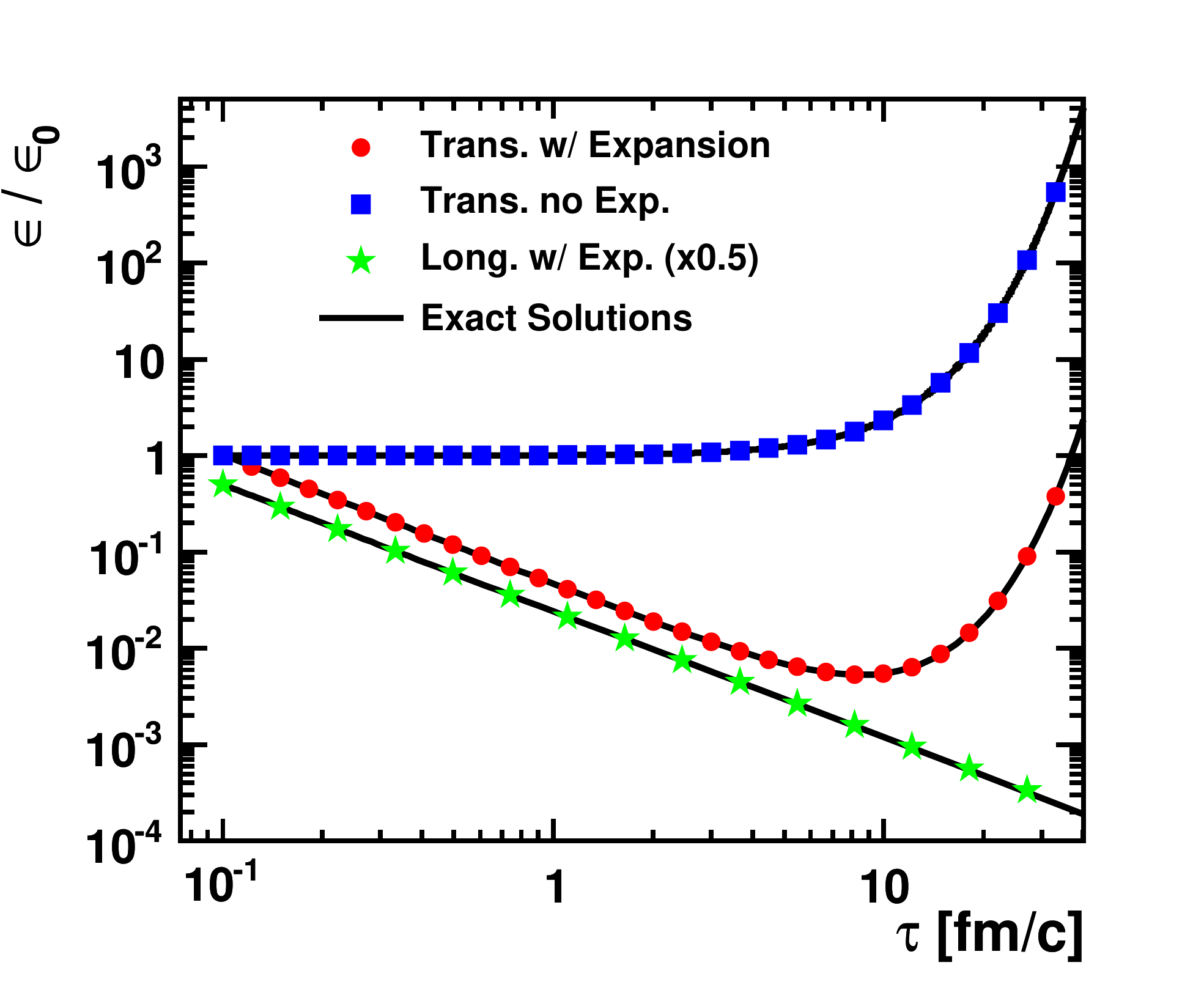}}
\caption{ (color online)
The ratio of the energy density to its initial value for a exponential energy density profile with no Bjorken expansion (blue squares), a transverse profile with Bjorken expansion (red stars),
and a longitudinal profile with Bjorken expansion (green triangles).
In each case the calculation produces the correct energy density.
}
\label{fig:expE}
\end{figure}

One can then solve for the evolution of the energy density, which can be written in terms of the velocity as
\begin{equation}
\frac{\dot{\epsilon}}{\epsilon} = \frac{u_0}{u^2(1-c_s^2) +1} \cdot \left[ \frac{(1-c_s^2)u}{R} - \frac{(1+c_s^2)u_0}{\tau} \right] \label{expHydro_EB}
\end{equation}
The general case can be easily integrated numerically, but there is a semi-analytic solution for the case with no Bjorken expansion.
One exploites a change of variables to arrive at an expression for the energy density given the velocity
\begin{eqnarray}
\int \frac{ d\epsilon}{\epsilon} &= & \int du \frac{dt}{du} \frac{d\epsilon}{dt}  \nonumber \\
\epsilon & = & \epsilon_0 \cdot exp \left[ \frac{1-c_s^4 }{2 c_s^2} u^2 \right] \label{expHydro_EU}
\end{eqnarray}
The results for a system with no expansion, a transverse system with expansion, and a longitudinal system with expansion are shown in Fig. \ref{fig:expE}. 

\end{document}